\documentclass[12pt]{article}
\def\reference{\parskip 0pt\par\noindent\hangindent 0.5 truecm}

\usepackage{psfig}
\usepackage{epsfig}
\usepackage{amssymb}

\newcommand{\beqn}{\begin{equation}}
\newcommand{\eeqn}{\end{equation}}

\newcommand{\Msun}{\ensuremath{M_{\odot}}}
\newcommand{\Lsun}{\ensuremath{L_{\odot}}}
\newcommand{\Mcmin}{\ensuremath{M_{\rm c}^{\rm min}}}
\newcommand{\McI}{\ensuremath{M_{\rm c}^1}}
\newcommand{\Menv}{\ensuremath{M_{\rm env}}}
\newcommand{\MO}{\ensuremath{M_{0}}}
\newcommand{\mdot}{\ensuremath{\dot{M}}}
\newcommand{\Lmax}{\ensuremath{\lambda_{\rm max}}}
\begin{document}
%
%
\title{Parameterising the third dredge-up in asymptotic giant branch stars}
%


\author{A.~I. Karakas, $^{1}$ 
 J.~C. Lattanzio, $^{1}$ \&
 O.~R. Pols $^{2,1}$ 
} 

\date{}
\maketitle

{\center
$^1$ School of Mathematical Sciences, Monash University, 
Wellington Rd, Clayton, Australia, 3800\\amanda.karakas@maths.monash.edu.au\\[3mm]
$^2$Astronomical Institute Utrecht, Postbus 80000, 3508 TA Utrecht, the Netherlands 
}

%
\begin{abstract}
We present new evolutionary sequences for low and
intermediate mass stars ($1\Msun$ to $6\Msun$) for three
different metallicities, $Z = 0.02,\,0.008$ and $0.004$.
We evolve the models from the pre-main sequence
to the thermally-pulsing asymptotic giant branch phase.  
We have two sequences of models for each mass, one which includes
mass loss and one without mass loss.
Typically 20 or more pulses have been followed for each model,
allowing us to calculate the third dredge-up 
parameter for each case.  Using the results from this 
large and homogeneous set of models, we present an
approximate fit for the core mass at the first thermal pulse, $\McI$, as 
well as for the third dredge-up efficiency parameter, $\lambda$, and the
core mass at the first dredge-up episode, $M_{\rm c}^{\rm min}$,
as a function of metallicity and total mass. We also examine the effect 
of a reduced envelope mass on the value of $\lambda$.


\end{abstract}

{\bf Keywords:}
stars: AGB and post-AGB -- stars: evolution -- stars: interiors -- stars: low mass

\bigskip

%
%

\section{Introduction}

The ascent of the asymptotic giant branch (AGB)
is the final nuclear-burning stage 
in the life of stars with masses between about $1$
and $8\Msun$. The combination of extensive nucleosynthesis and high
mass loss makes these stars crucial for understanding the chemical composition
of galaxies. For recent reviews see
Iben (1991), Frost \& Lattanzio (1995) and Busso, Gallino \& Wasserburg (1999).

\begin{figure}[t]
\begin{center}
\psfig{file=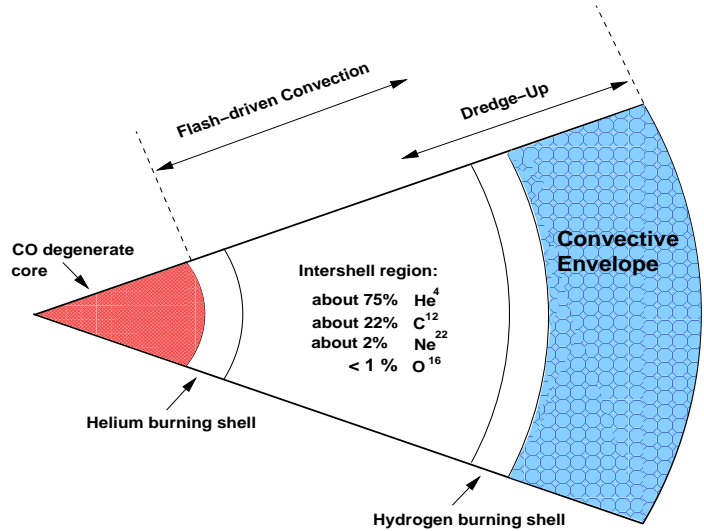,width=9.4cm,height=7.0cm,angle=0} 
\caption{Schematic structure of an AGB star, showing the degenerate CO core surrounded by 
a He-burning shell above the core and a H-burning shell below the deep convective envelope.
The burning shells are separated by an intershell region rich in helium ($\sim 75\%$)
and carbon ($\sim 22\%$) with some oxygen.  
Note this diagram is not to scale. The ratio of the radial
thickness of the H-exhausted core compared to the envelope is about $1 \times 10^{-5}$.}
\label{fig:agbStruct} 
\end{center}           
\end{figure}

Very briefly, an AGB star is characterized by two nuclear burning shells, 
one burning helium (He) above a degenerate carbon-oxygen core and 
another burning hydrogen (H), below
a deep convective envelope as shown in Figure~\ref{fig:agbStruct}.  
The He-burning shell is thermally unstable, and 
pulses every $10^{4}\,$years or so, depending on the
core mass\footnote{unless otherwise specified, by ``core'' 
we mean the H-exhausted core} and composition of the star. 
In each thermal pulse (TP), the He-burning
luminosity can reach up to $L_{\rm He} \sim 10^{8}\Lsun$, most of 
which goes into expanding the outer layers.
This strong expansion drives the H-shell 
to cooler, less dense regions which has the effect of 
extinguishing the H-shell. 
The inner edge of the  deep convective envelope can then 
move inward (in mass) and mix to the surface the products of 
internal nucleosynthesis.
This mixing event, which can occur periodically (after each TP),
is known as the third dredge-up (TDU) and is the mechanism
for producing (single) carbon stars.
Following dredge-up, the star contracts, re-igniting the 
H-shell and enters a phase of 
quiescent H-burning, known as the interpulse phase. 
The thermally pulsing AGB (TP-AGB) is defined as the phase 
after the first thermal pulse to the time when 
the star ejects its envelope, terminating the AGB phase.

The efficiency of the TDU is quantified by the parameter $\lambda$, 
which is the ratio of mass dredged up by the convective envelope, 
$\Delta M_{\rm dredge}$, 
to the amount by which the core mass increased due to H-burning during the 
preceding interpulse period, $\Delta M_{c}$,  
\beqn
 \lambda  = \frac{\Delta M_{\rm dredge}} {\Delta M_{c}}.  \label{eq:lambda}
\eeqn
The value of $\lambda$ 
depends on physical parameters such as the core mass, 
metallicity (and hence opacity)
as well as the total mass of the star.
Exactly how $\lambda$ depends on these quantities is still
unknown. The two main reasons for this are the difficulty in
locating the inner edge of the convective envelope during the
dredge-up phase (Frost \& Lattanzio 1996,
Mowlavi 1999) and the huge computer resources required to explore
an appropriate range of mass and composition over such a
computationally demanding evolutionary phase. Without a systematic
investigation of the dredge-up law, only certain trends have been
identified by extant models, such as the increase of $\lambda$ with
decreasing $Z$ and increasing mass (Boothroyd \& Sackmann 1988),
and the fact that below some critical envelope mass, the dredge-up 
ceases altogether (Straniero et al. 1997).

The convenient fact that the stellar luminosity on the AGB is a nearly 
linear function of the H-exhausted core mass has stimulated the 
development of ``synthetic'' AGB evolution models, as a quick way of
simulating stellar populations on the AGB. The main observational
constraint which models must face is the carbon star luminosity
function (CSLF) for the Magellanic Clouds. In some
synthetic AGB evolution calculations, 
e.g. as performed by Groenewegen \& de Jong (1993) and Marigo (1996),
$\lambda$ is treated as a constant free parameter,  calibrated by
comparison with the CSLF.

Synthetic codes enable us to investigate a diverse 
range of problems, such as binary
population synthesis (Hurley, Tout \& Pols, 2002), 
AGB population studies (Groenewegen \& de Jong 1993),
and the calculation of stellar yields from AGB stars 
(Marigo 1996, 1998a,b, 2001; van~den~Hoek \&
Groenewegen 1997).
Most parameterisations used in 
synthetic evolution studies are found either empirically from 
observations (such as mass loss)
or from results from full stellar calculations, such
as the core-mass-interpulse-period relation.  Currently there 
are no parameterisations in the literature
based on detailed evolutionary models
that describes the behaviour of $\lambda$ with total mass, 
metallicity, age and/or core mass, for the reasons given above.

With current computing power the problem becomes time 
consuming rather than impossible. Hence we have embarked on just such
an exploration of relevant parameter space using full detailed
evolutionary models. Our aim is to determine the dependence of evolutionary
behaviour (such as the dredge-up law) on the various stellar
parameters, and to provide these in a form suitable for use in
synthetic population studies.

This paper is organised as follows. First we discuss the stellar 
evolution models and the code used to calculate them. In the second 
section we discuss our method for parameterising
dredge-up and give the fitting formulae we found from the stellar models to 
describe the core mass at the first thermal pulse, $\McI$, the core mass at
the first TDU episode, $\Mcmin$, and 
$\lambda$ as functions of initial mass 
and metallicity.  We finish with a discussion.

\section{Stellar Models}  \label{section:models}
Evolutionary calculations were performed with the Monash version of the
Mt Stromlo Stellar Evolution Code (Frost, 1997; Wood \& Zarro, 1981) 
updated to include the OPAL opacity tables of Iglesias \& Rogers
(1996).  We ran about 60 sequences of stellar models, from the zero-age
main-sequence (ZAMS) to near the end of the TP-AGB 
for three different compositions: $Z = 0.02, 0.008$ and $0.004$.
For each composition we cover a range in mass between $1$ and
$6\Msun$. We do not include overshooting in the convective cores 
of intermediate mass stars  during H burning on the main-sequence, 
although there is observational evidence for a small overshoot region.

\subsection{Convection and  Dredge-up} \label{section:convdu}
The amount of third dredge-up found in evolutionary calculations 
crucially depends 
on the numerical treatment of convective boundaries:~many codes do 
not find any dredge-up for low-mass stars without some form 
of overshoot (Mowlavi 1999; Herwig 1997).  
Herwig (2000) found
very efficient dredge-up, with $\lambda > 1$, in a $3\Msun$ $Z=0.02$
model with diffusive convective
overshoot on all convective boundaries but no dredge-up for the same 
mass without overshoot.
Pols \& Tout (2001) found very efficient dredge-up,
with $\lambda \sim 1$, in a $5\Msun$ $Z=0.02$ model using a completely 
implicit and simultaneous 
solution for stellar structure, nuclear burning and convective mixing.
Frost \& Lattanzio (1996) found the treatment of entropy to affect 
the efficiency of dredge-up 
and Straniero et al. (1997) found the space and time resolution 
to be important.

In view of this strong dependence on numerical details, it is important 
to specify carefully how we treat convection.
We use the standard mixing-length theory for convective regions, with
a mixing-length parameter $\alpha = l / H_P = 1.75$, and
determine the border by applying the Schwarzschild criterion.
Hence we do not include convective overshoot, in the usual sense.
We do, however, recognize the discontinuity in the ratio $r$ of the
radiative to adiabatic temperature gradients at the bottom edge of
the convective envelope during the dredge-up phase. We search for a 
neutral border to the convective zone, in the manner described in
Frost \& Lattanzio (1996). Briefly, we extrapolate (linearly, in mass)
the ratio $r$ from
the last convective point to the first radiative point, and if $r>1$
then we include this point in the convective region for the next
iteration on the structure. We remind the reader that this algorithm
sometimes fails, in the sense that the convective envelope grows
deeper and then retreats, with succeeding iterations. In such a case, 
we take the deepest extent as the {\it mixed\/} region, even if the
{\it convective\/} region is shallower when the model converges.

Finally we note that, although we believe our treatment of convection is
reasonable, our results cannot be regarded as the definitive solution
to the difficult problem of third dredge-up. However, the important
point is that all our models are computed using the same algorithm. 
Together they constitute, for the first time, an internally consistent set
of models covering a wide range in mass and metallicity. 

\subsection{Mass Loss} \label{section:massloss}
Mass loss is a crucial part of AGB evolution, and seriously affects
dredge-up in two ways. Firstly, for the more massive stars 
dredge-up can be terminated when the
envelope mass decreases below some critical value.  
Secondly, for lower masses,
mass loss may terminate the AGB evolution before the H-exhausted core
reaches the minimum value for dredge-up to occur. 
However, the mass-loss rate in  AGB stars is very uncertain, and 
for this reason we calculate each stellar sequence both with and
without mass loss. By neglecting mass loss, we find the limiting behaviour
of dredge-up for each model we calculate. In Section~\ref{section:param}
we parameterise this dredge-up behaviour in the absence of mass loss.
When this parameterisation is 
used in synthetic evolutionary
calculations, the chosen mass-loss law will determine if the models
reach the limiting behaviour we provide.
The subsequent AGB evolution and 
dredge-up will then be modified by the choice of mass-loss law. For
example, we will determine a minimum core mass for dredge-up at a given 
mass and composition in the case without mass loss, and whether the
model reaches this core mass or not will depend on the chosen mass-loss 
rate. Alternatively, a particular mass-loss law may or may not prevent 
a model from reaching the asymptotic value for $\lambda$, which can
only be determined from full stellar models
without the inclusion of mass loss.

We also ran one set of models with our preferred mass-loss law.
We use the Reimers (1975) formula on the red giant branch 
with $\eta = 0.4$ and then the prescription of Vassiliadis \& Wood
(1993) (hereafter VW93) on the AGB.
VW93 parameterised the mass-loss rate as a function of pulsation period, 
\beqn
  \log \left( \frac{dM}{dt} \right) = - 11.4 + 0.0125 P,  \label{eq:vw93}
\eeqn
where the mass-loss rate is in $\Msun \, {\rm yr}^{-1}$ and 
$P$ is the pulsation period in days, given by
\beqn
  \log P = - 2.07 + 1.94\log R -0.9\log M,
\eeqn
where $R$ and $M$ are the stellar radius and mass in solar units.
For $P \geqslant 500\,$days, the mass-loss rate given in 
Equation~\ref{eq:vw93} is truncated at
\beqn
   \frac{d M}{dt} = \frac {L}{c \, v_{\rm exp}},
\eeqn
corresponding to a radiation-pressure driven wind ($L$ is the stellar
luminosity in solar units).  The 
wind expansion velocity, $v_{\rm exp}$ is also 
taken from VW93 and is given by
\beqn
  v_{\rm exp} = -13.5 + 0.056 P,
\eeqn
where $v_{\rm exp}$ is in ${\rm km}\,{\rm s}^{-1}$, and is limited to
a maximum of 15~km\,s$^{-1}$.

\subsection{Evolutionary Sequences} \label{section:seqs}
The models were evolved from the main-sequence,
through all intermediate stages, including the core-helium flash for
initial mass $\MO \lesssim 2.5\Msun$.
Most models without mass loss were evolved
until $\lambda$ reached an asymptotic value. 
Models with mass loss were
evolved until convergence difficulties ceased the calculation, which
was very near the end of the AGB phase.
Typically the final envelope mass was quite small, 
$\Menv \lesssim 0.1$, for low-mass models ($\MO \lesssim 2.5$) 
and $\Menv \sim 1\Msun$ for
intermediate mass stars ($\MO \gtrsim 3$). The remaining evolution is
extremely brief, because the mass-loss rate is so high (typically
a few times 10$^{-5} \Msun\,$yr$^{-1}$) at this stage.

Evolutionary sequences were calculated for stars with 
masses:\footnote{Note all masses quoted
are the ZAMS initial mass, in solar units, unless otherwise stated.} 
$1$, $1.25$, $1.5$, $1.75$, $1.9$, $2$,
$2.1$, $2.25$, $2.5$, $3$, $3.5$, $4$, $5$ and $6\Msun$.
The initial compositions used are shown in 
Table~\ref{tab:comp} and are similar to solar composition,
Large Magellanic Cloud (LMC) and Small Magellanic Cloud (SMC) 
composition and were chosen to be
consistent with the models of Frost (1997).  
\begin{table}
\centering
\caption{Initial compositions (in mass-fractions) used for stellar models:}
\vspace{2mm}
\begin{tabular}{@{}cccc@{}}  \hline
             & $Z = 0.02$   &     $Z = 0.008$  &    $Z = 0.004$   \\ 
             &  solar       &     LMC          &    SMC           \\ \hline
   X         &    0.6872    &    0.7369        &    0.7484       \\
   Y         &    0.2928    &    0.2551        &    0.2476       \\
   $^{12}$C         &  2.9259(-3)  &  9.6959(-4)      &  4.8229(-4)      \\                          
   $^{14}$N         &  8.9786(-4)  &  1.4240(-4)      &  4.4695(-5)      \\
   $^{16}$O         &  8.1508(-3)  &  2.6395(-3)      &  1.2830(-3)       \\
Other $Z$    &  8.0253(-3)  &  4.2484(-3)      &  2.1899(-3)       \\ \hline
\end{tabular} \label{tab:comp}
\end{table}

\subsection{Model Results} \label{section:results}

We found that the third dredge-up behaviour of models
that experienced the second dredge-up (SDU) 
(generally masses $\MO \gtrsim 4$ depending on $Z$, 
or core masses greater than $0.8\Msun$), differs qualitatively
from that of lower-mass models.
To find the minimum mass for the SDU for the three different 
compositions we ran a few models to the start of the TP-AGB only.   
We found the SDU at $M \geqslant 4.05\Msun$ for $Z=0.02$,
$M \geqslant 3.8\Msun$ for $Z=0.008$ and 
$M \geqslant 3.5\Msun$ for $Z=0.004$. 

\begin{figure}
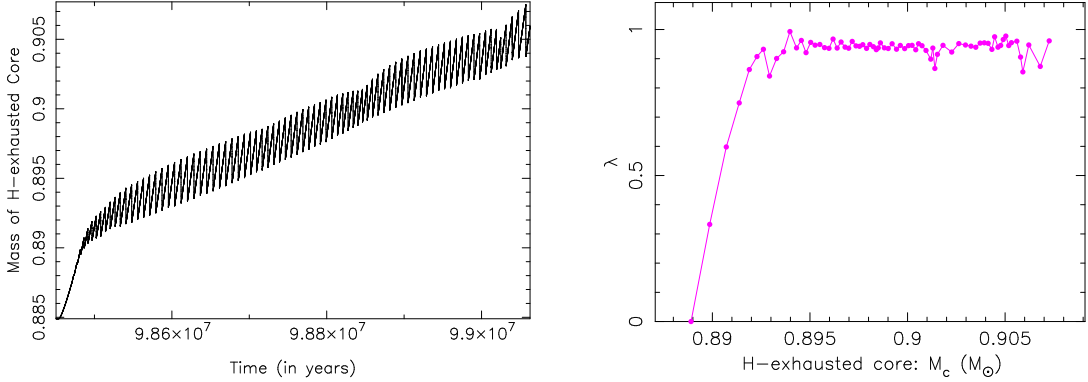

\begin{minipage}{0.5\textwidth}
\psfig{file=fig2.eps,width=5.0cm,angle=270}
\end{minipage}
\hspace{\fill}
\begin{minipage}{0.5\textwidth}
\psfig{file=fig3.eps,width=5.0cm,angle=270}
\end{minipage}
\caption{({\it Left}) H-exhausted core mass 
$M_{\rm c}$ against time (years) for the $5\Msun$, $Z = 0.004$ model 
with mass loss. This calculation covered
74 pulses, which equated to roughly $7 \times 10^{5}$ individual 
stellar models.
({\it Right}) The dredge-up parameter $\lambda$ against core mass. Dredge-up 
increased very quickly with pulse number, reaching $0.9$ within four
TPs. }
 \label{fig:m5z004}
\end{figure}

As an example of our results for higher masses, we show 
the $5\Msun$, $Z=0.004$ model with mass loss in 
Figure~\ref{fig:m5z004}. This sequence shows
74 TPs with the last calculated model having $\Menv = 0.944$ 
and a core mass of $M_{\rm c} = 0.906$.  
The dredge-up parameter
$\lambda$ is seen to increase very quickly, reaching a value near $0.96$ in
only four pulses and maintaining that value until the end of the
calculation.  We see that $\lambda$ oscillates
a little near the end, between $0.85$ and $0.96$; this may indicate
the imprecision of the dredge-up algorithm.
We find that in all our higher-mass models 
$\lambda$ reached an asymptotic value of about
$0.9$ or higher, regardless of composition and mass loss.

Moving to the lower mass models, we first compare those
with and without mass loss. Models with mass loss
have shallower dredge-up, sometimes none at all, compared to the
limiting values found with constant mass.
We note that many of the models with mass loss
do not reach the minimum core mass for TDU, $\Mcmin$.
We will therefore parameterize $\Mcmin$ from models without mass loss. 

\begin{figure}
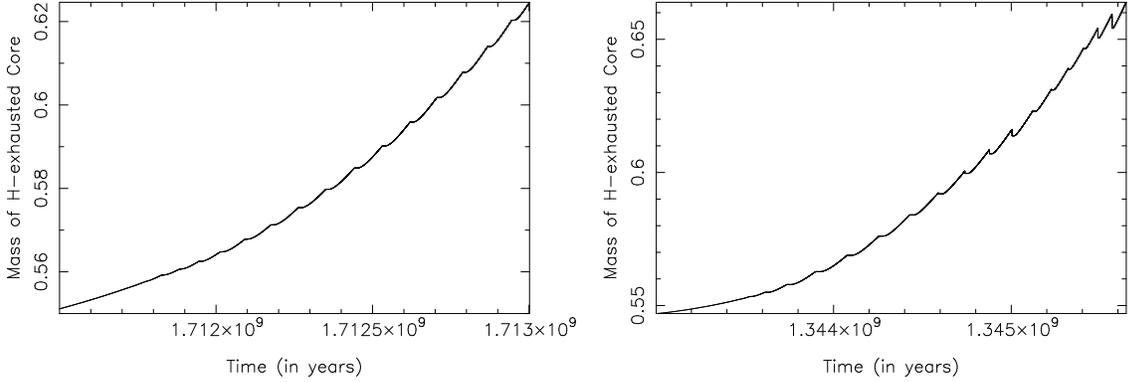

\begin{minipage}{0.5\textwidth}
\psfig{file=fig4.eps,width=5.0cm,angle=270}
\end{minipage}
\hspace{\fill}
\begin{minipage}{0.5\textwidth}
\psfig{file=fig5.eps,width=5.0cm,angle=270}
\end{minipage}

\caption{({\it Left}) $M_{\rm c}$ against time (years) for a $\MO = 1.75\,\Msun$, 
$Z=0.02$ model with mass loss. No dredge-up was found. The final $\Menv$ for 
this model was $0.0087\,\Msun$.  
({\it Right}) $M_{\rm c}$ against time (years) for a $\MO = 1.75\,\Msun$, $Z=0.004$
model with mass loss.
This model experiences appreciable though erratic dredge-up. $\lambda$ quickly 
reached $0.26$, 
before being reduced back to zero, then increased right at the end to reach $0.6$. 
This star became a carbon star when
$M_{\rm c} = 0.615\Msun$, $L \sim 9700 \Lsun$, and $M_{\rm bol} = - 5.43$.
The final dredge-up episode takes place
with an $\Menv = 0.175\,\Msun$ after the last calculated
TP. The final $\Menv = 0.025\,\Msun$.}
\label{fig:lowmass} 
\end{figure}

As with previous calculations (Boothroyd \& Sackmann 1988; 
Vassiliadis 1992; Straniero 1997)
we found that $\lambda$ increases with decreasing metallicity
for a given mass (with or without mass loss) for the lower-mass models.
For example  we found no 
dredge-up for a $1.75\Msun,\,Z=0.02$ model with mass loss 
but the $1.75\Msun,\,Z=0.004$ model with mass loss became a 
carbon star, with a  maximum $\lambda \sim 0.6$, 
as shown in Figure~\ref{fig:lowmass}.  

One of the aims of the models with mass loss was to examine how 
$\lambda$ decreases with decreasing envelope mass, $\Menv$, and the 
critical envelope mass for which dredge-up ceases. Unfortunately, 
the higher mass models ($M \gtrsim3$) suffered convergence problems
before reaching this critical envelope mass. 
We find no systematic decrease of $\lambda$ as the envelope mass
decreases (see Fig.~\ref{fig:m5z004}).
For $Z=0.004$ and $Z=0.008$, the low-mass models that do experience
dredge-up have $\lambda > 0$ as long as  $\Menv \gtrsim 0.2$, which
is thus our estimate of the critical envelope mass for TDU to occur.

Table~\ref{tab:z02} presents results for the $Z=0.02$ models with ($\mdot
\neq 0$) and without mass loss ($\mdot = 0$). 
The first column shows the initial mass ($\MO$), and the zero-age horizontal
branch (ZAHB) mass in
parentheses for low-mass stars. The second column gives the core mass 
at the first thermal pulse, $\McI$, column three gives $\Lmax$, the 
maximum $\lambda$ for that model, column four 
the core mass at the first dredge-up episode, $\Mcmin$ and column 
five the number of thermal pulses calculated.
Some low-mass models $M \leqslant 3\Msun$ (depending on $Z$) 
do not undergo enough thermal pulses with 
dredge-up to obtain an asymptotic value. In these cases we give the 
largest value found for $\lambda$, denoted by ``L'' as the value 
of $\Lmax$.
We find no dredge-up for $Z=0.02$ mass-loss models of low-mass, $\MO \leqslant
2\Msun$. Between $2 < \MO/\Msun < 3$, we find $\lambda$  to be smaller
for models with mass loss than for those without.  There is
no appreciable difference in the values of $\Lmax$ and $\Mcmin$ 
for the $\MO \geqslant 3\Msun$ 
models with or without mass loss.

Table~\ref{tab:z008} presents results for $Z=0.008$.  
For these stars with an LMC composition,
the effect of mass loss is seen at lower masses, 
with masses below $1.5\Msun$ being the
most strongly affected.   By $1.9\Msun$, mass loss has little 
effect on the depth of dredge-up, 
where we find $\Lmax = 0.5$ for the model with mass loss 
compared with $\Lmax = 0.6$ for the model without mass loss.  
Note that for sequences without mass loss we terminated the
calculation once an asymptotic value of $\lambda$ was found.
Table~\ref{tab:z004} presents results for $Z=0.004$.  
For this composition mass loss only affects models with $M < 1.5\Msun$
and both the $1.5$ and $1.75\Msun$ models with mass loss became carbon stars.  

\begin{figure}[t]
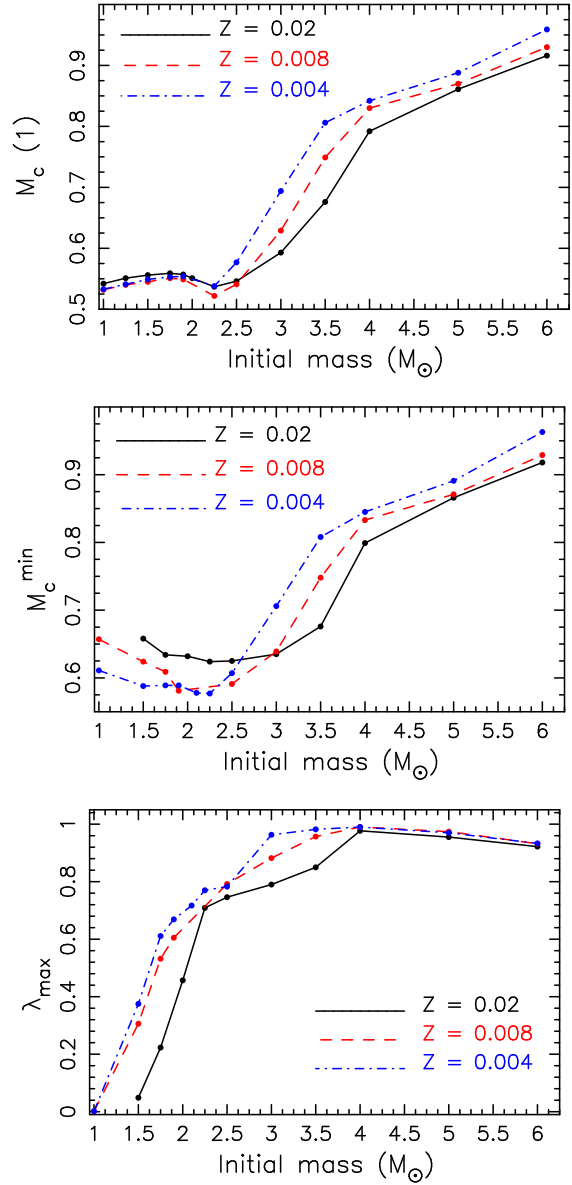

\begin{tabular}{c}
  \psfig{file=fig6.eps,width=5.0cm,angle=270} \\
   \psfig{file=fig7.eps,width=5.0cm,angle=270} \\
  \psfig{file=fig8.eps,width=5.0cm,angle=270}
\end{tabular}
\\ 
\caption{The $\McI$, $\Mcmin$ and $\Lmax$ plotted against initial
mass for the $Z=0.02$, $Z=0.008$ and $Z=0.004$ models without mass loss.
In each panel, the blue solid line refers to the $Z=0.02$ models, the 
black dashed line to the $Z=0.008$ models and the red dash-dotted line to the 
$Z=0.004$ models.} 
\label{fig:lmax-mcmin} 
\end{figure}

Figure~\ref{fig:lmax-mcmin} shows the $\McI$, $\Mcmin$ and $\Lmax$ values 
from Tables~\ref{tab:z02},~\ref{tab:z008} and~\ref{tab:z004} plotted 
against the initial mass, for all compositions calculated without mass 
loss.
From Tables~\ref{tab:z02},~\ref{tab:z008} 
and~\ref{tab:z004} we find that $\McI$ and $\Mcmin$ are largely independent 
of mass loss.
The behaviour of $\McI$ in Figure~\ref{fig:lmax-mcmin} is similar for 
low-mass stars independent of $Z$,
with $\McI \sim 0.55$. There is a dip in $\McI$ at $\MO \approx 2.25\Msun$, 
corresponding to the transition
from degenerate to non-degenerate He-ignition,
followed by an increase with increasing initial mass.  
For models undergoing the SDU ($\MO \gtrsim 4\Msun$) the variation is 
nearly linear. 
The value of $\Mcmin$ for low-mass stars ($\MO \lesssim 2.5\Msun$) decreases 
somewhat with increasing mass and decreasing $Z$, and then shows a similar 
increase with mass as does $\McI$.  For $\MO \gtrsim 4\Msun$, $\McI$ and
$\Mcmin$ are nearly equal, i.e.\ dredge-up sets in almost immediately
after the first pulse. 

A comparison with current synthetic calculations is useful. Most
calculations have so far assumed a constant value of $\Mcmin$ 
(Groenewegen \& de Jong, 1993), but Marigo (1998a) attempted
to improve on this. She assumed dredge-up to occur if, following a pulse,  
the temperature at the base of the
convective envelope reached a specified value $T_{\rm b}^{\rm dred}$.
We compared our $Z=0.008$ results for $\Mcmin$ with Figure~3 in Marigo
(1998a). For $M \leqslant 2.5$, our
values for $\Mcmin$ agree well with her values for 
$\log T_{\rm b}^{\rm dred} = 6.7$.
Indeed,  we find $\log T_{\rm b}^{\rm dred} = 6.7 \pm 0.2$ for 
all our low-mass models ($M \leqslant 2.5\Msun$) but showing a slight $Z$ 
dependence.  The $Z=0.02$ models required slightly higher temperatures
than the lower-metallicity models, with  
$\log T_{\rm b}^{\rm dred} = 6.8 \pm 0.1$, 
whilst the $Z=0.008$ and $Z=0.004$ models required
$\log T_{\rm b}^{\rm dred} = 6.6 \pm 0.1$ for dredge-up.  
We also note that for deep dredge-up 
($\lambda \gtrapprox  0.5$) the temperature must be higher, 
$\log T \approx 6.9$.

\begin{table}
\begin{center}
\caption{$\McI$, $\Lmax$ and $\Mcmin$ for $Z=0.02$ models without mass loss
($\mdot = 0$) and for models with mass loss ($\mdot \ne 0$). 
Column one gives the initial mass, with ZAHB mass in
parentheses if applicable, column two $\McI$,  
column three $\Lmax$, column four $\Mcmin$ and column five 
the number of TPs. Blank entries in the table reflect masses that were not
calculated. The meaning of the symbol `L' is explained
in the text.}
\vspace{4mm}
\begin{tabular}{@{}|c|c|c|c|c|c|c|c|c|@{}}  \hline 
$\MO$ & \multicolumn{2}{c|}{$M_{\rm c}^{1}$} & \multicolumn{2}{c|}{$\Lmax$}  
& \multicolumn{2}{c|}{$\Mcmin$} &
\multicolumn{2}{c|}{No. of TPs} \\ \hline \hline
     & \small{$\mdot = 0$} & \small{$\mdot \ne 0$}  & \small{$\mdot = 0$} & \small{$\mdot \ne 0$} & 
\small{$\mdot = 0$} & \small{$\mdot \ne 0$} & \small{$\mdot = 0$} & \small{$\mdot \ne 0$} \\ \hline
1.0  (0.82) &       &  0.542 &        &  0.0     &        &  --   &    & 11  \\
1.25 (1.13) & 0.556 &  0.551 & 0.0    &  0.0     &  --    &  --   & 24 &  10   \\
1.5 (1.41)  & 0.560 &  0.556 & 0.0486(L) &  0.0  &  0.658 &  --   & 24 &  13   \\
1.75 (1.68) & 0.561 &  0.559 & 0.223  & $0.0$ &  0.634 &  --   & 28 &  15   \\
1.9 (1.84)  &       &  0.557 &        & $0.0$ &        &  --   &    &  18   \\
2.0 (1.96)  & 0.554 &  0.551 & 0.457(L) & 0.00145(L) &  0.632 & 0.633 & 27 &  21   \\
2.25        & 0.540 &  0.537 & 0.709  & 0.305(L) &  0.624 & 0.620 & 37 &  28   \\
2.5         & 0.549 &  0.546 & 0.746  & 0.538(L) &  0.625 & 0.623 & 36 &  30   \\ 
3.0         & 0.595 &  0.593 & 0.790  & 0.805    &  0.635 & 0.630 & 25 &  25   \\
3.5         & 0.662 &  0.676 & 0.850  & 0.880    &  0.676 & 0.690 & 26 &  22   \\ 
4.0         & 0.793 &  0.792 & 0.977  & 0.958    &  0.799 & 0.797 & 22 &  17   \\
5.0         & 0.862 &  0.861 & 0.955  & 0.957    &  0.866 & 0.864 & 28 &  25   \\
6.0         & 0.915 &  0.916 & 0.922  & 0.953    &  0.918 & 0.919 & 65 &  40   \\ \hline
\end{tabular} \label{tab:z02}
\end{center}
\end{table}

\begin{table}
\begin{center}
\caption{Table of $\McI$, $\Lmax$ and $\Mcmin$ for $Z=0.008$.}
\vspace{4mm}
\begin{tabular}{@{}|c|c|c|c|c|c|c|c|c|@{}}  \hline
$\MO$ & \multicolumn{2}{c|}{$\McI$} & \multicolumn{2}{c|}{$\Lmax$}  
& \multicolumn{2}{c|}{$\Mcmin$} &
\multicolumn{2}{c|}{No. of TPs} \\ \hline \hline
     & \small{$\mdot = 0$} & \small{$\mdot \ne 0$}  & \small{$\mdot = 0$} & \small{$\mdot \ne 0$} & 
\small{$\mdot = 0$} & \small{$\mdot \ne 0$} & \small{$\mdot = 0$} & \small{$\mdot \ne 0$} \\ \hline
1.0  (0.85) & 0.535 & 0.532 & 0.0016(L) & 0     &  0.657 & --    & 22 &  11    \\
1.25 (1.14) &       & 0.540 &        & 0        &        & --    &    &  12   \\
1.5  (1.42) & 0.550 & 0.545 & 0.306  & 0.0842(L) &  0.624 & 0.610 & 21 &  15   \\
1.75 (1.68) & 0.555 & 0.551 & 0.532(L) & 0.325(L) &  0.609 & 0.595 & 21 &  15    \\
1.9  (1.85) & 0.551 & 0.549 & 0.605(L) & 0.500(L) &  0.581 & 0.594 & 21 &  18   \\
2.1         & 0.540 &       & 0.656  &          &  0.596 &       & 22 &       \\
2.25        &       & 0.522 &        &  0.727(L) &        & 0.585 &    &  27   \\  
2.5  	    & 0.540 & 0.541 & 0.792(L)  &  0.805   &  0.591 & 0.587 & 27 &  28  \\ 
3.0         & 0.629 & 0.629 & 0.882  &  0.897   &  0.639 & 0.648 & 20 &  29   \\
3.5  	    & 0.744 & 0.749 & 0.957  &  0.980   &  0.748 & 0.756 & 22 &  21  \\
4.0         & 0.830 & 0.830 & 0.990  &  0.970   & 0.833  & 0.833 & 17 &  24   \\
5.0         & 0.870 & 0.870 & 0.974  &  0.980   & 0.871  & 0.872 & 27 &  58 \\ 
6.0         & 0.926 & 0.930 & 0.932  &  0.947   & 0.929  & 0.933 & 26 &  68   \\ \hline
\end{tabular} \label{tab:z008}
\end{center}
\end{table}

\begin{table}
\begin{center}
\caption{Table of $\McI$, $\Lmax$ and $\Mcmin$ for $Z=0.004$.}
\vspace{4mm}
\begin{tabular}{@{}|c|c|c|c|c|c|c|c|c|@{}}  \hline
$\MO$ & \multicolumn{2}{c|}{$\McI$} & \multicolumn{2}{c|}{$\Lmax$}  & \multicolumn{2}{c|}{$\Mcmin$} &
\multicolumn{2}{c|}{No. of TPs} \\ \hline \hline
     & \small{$\mdot = 0$} & \small{$\mdot \ne 0$}  & \small{$\mdot = 0$} & \small{$\mdot \ne 0$} & 
\small{$\mdot = 0$} & \small{$\mdot \ne 0$} & \small{$\mdot = 0$} & \small{$\mdot \ne 0$} \\ \hline
1.0  (0.87) & 0.541 & 0.533 & 0      & 0.003(L)  &  --  & 0.611 & 22  & 14      \\
1.25 (1.16) &       & 0.541 &        & 0.0787(L) &      & 0.600 &     & 14    \\
1.5  (1.43) & 0.551 & 0.549 & 0.375(L) & 0.325(L) & 0.588 & 0.601 & 15  & 15    \\
1.75 (1.70) & 0.558 & 0.553 & 0.611(L) & 0.593(L) & 0.589 & 0.592 & 16  &  18    \\
1.9  (1.86) & 0.558 & 0.554 & 0.669  & 0.612(L) & 0.589 & 0.593 & 18  &  18   \\
2.1         & 0.550 &       & 0.717(L) &          & 0.578 &       & 16  &       \\
2.25        & 0.537 & 0.538 & 0.770  & 0.767    & 0.577 & 0.577 & 26  &  26   \\
2.5         & 0.578 & 0.577 & 0.783  & 0.832    & 0.607 & 0.603 & 15  &  28  \\ 
3.0         & 0.699 & 0.694 & 0.963  &  0.952   & 0.706 & 0.702 & 16  &  26   \\
3.5         & 0.804 & 0.806 & 0.982  &  0.998   & 0.808 & 0.809 & 20  &  23  \\
4.0         & 0.842 & 0.842 & 0.990  &  0.975   & 0.845 & 0.845 & 20  &  30   \\
5.0         & 0.889 & 0.888 & 0.970  & 0.960    & 0.891 & 0.890 & 24  & 74   \\
6.0         & 0.962 & 0.959 & 0.933  & 0.940    & 0.963 & 0.961 & 30  &  95  \\ \hline
\end{tabular} \label{tab:z004}
\end{center}
\end{table}

\section{Parameterising the Third Dredge-Up} \label{section:param}
First we will describe the fit we made to $\McI$ and then
$\Mcmin$ and $\Lmax$, followed by a simple prescription to 
model the variation of $\lambda$ with pulse number.

\subsection{The Fitting Formula for $\McI$}\label{section:param-mci}

Wagenhuber \& Groenewegen (1998) have provided a fitting formula for
the core mass at the first thermal pulse, $\McI$ as a function of mass
and metallicity (their equation 13). We have compared their Population I fit 
to our results for $Z=0.02$, and find qualitative agreement in the shape
of the formula but significant quantitative differences. The same is true for
lower metallicities, when we linearly interpolate the 
coefficients given in Wagenhuber \& Groenewegen (1998) for $Z = 0.008$ and 
$Z=0.004$ and compare the resulting relation to our models.

Here we provide modified coefficients for the fitting formula 
given by Wagenhuber \& Groenewegen (1998), instead of providing a completely
new fit to $\McI$ as we do for $\Mcmin$ and $\Lmax$. 
We choose to do this for two reasons.  Firstly, the shape of the
function provided by Wagenhuber \& Groenewegen (1998) for $\McI$ (equation 13a-c) 
is a very good approximation to the shape of the $\McI$-initial mass relation
we find from our models. 
Secondly, researchers who already use the Wagenhuber \& Groenewegen (1998) 
$\McI$ fit for Pop I and II stars in their synthetic evolution codes can 
easily convert to our fit for Pop I, LMC and SMC models. 
The modified coefficients to the Wagenhuber \& Groenewegen (1998) formula
can be found in the Appendix.
Figures~\ref{fig:mc1fit} shows the modified fits 
to $\McI$ plotted against the results from the models without mass loss.  
Note that the lines between $6\Msun$ and $8\Msun$ are
extrapolations from the fitting 
functions (valid to $6\Msun$) and may not reflect real model
behaviour,
although one test calculation was made for $M=6.5$ and $Z=0.02$ 
and did agree with the fit.

\begin{figure}
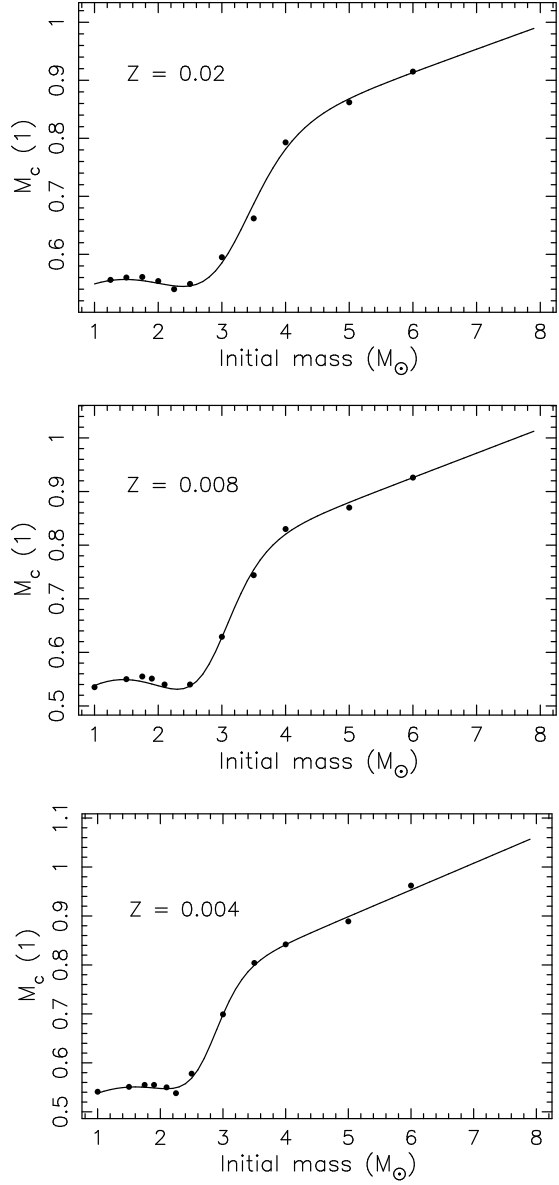

\begin{minipage}{0.5\textwidth}
\begin{tabular}{c} 
   \psfig{file=fig9.eps,width=5.0cm,angle=270}  \\
   \psfig{file=fig10.eps,width=5.0cm,angle=270} \\
   \psfig{file=fig11.eps, width=5.0cm,angle=270} \\
\end{tabular}
\end{minipage}
 \\
\caption{The fit to $\McI$ using the Wagenhuber \& Groenewegen (1998)
fit with modified coefficients (solid line) plotted with model results
for the $Z=0.02$, $Z=0.008$ and $Z=0.004$ models without mass loss.} 
\label{fig:mc1fit}
\end{figure}

\subsection{The Fitting Formulae for $\Mcmin$ and $\Lmax$}
We fit $\Lmax$ as a function of total mass
by using a rational polynomial and $\Mcmin$ by using a 
third order polynomial at low masses. At higher masses $\Mcmin$ simply
follows $\McI$.
We provide a separate fit 
for each composition, but interpolation between the coefficients 
of the polynomials should 
be possible for arbitrary $Z$ in the range $0.02$ to $0.004$. 

\begin{figure}
\begin{tabular}{c} 
   \psfig{file=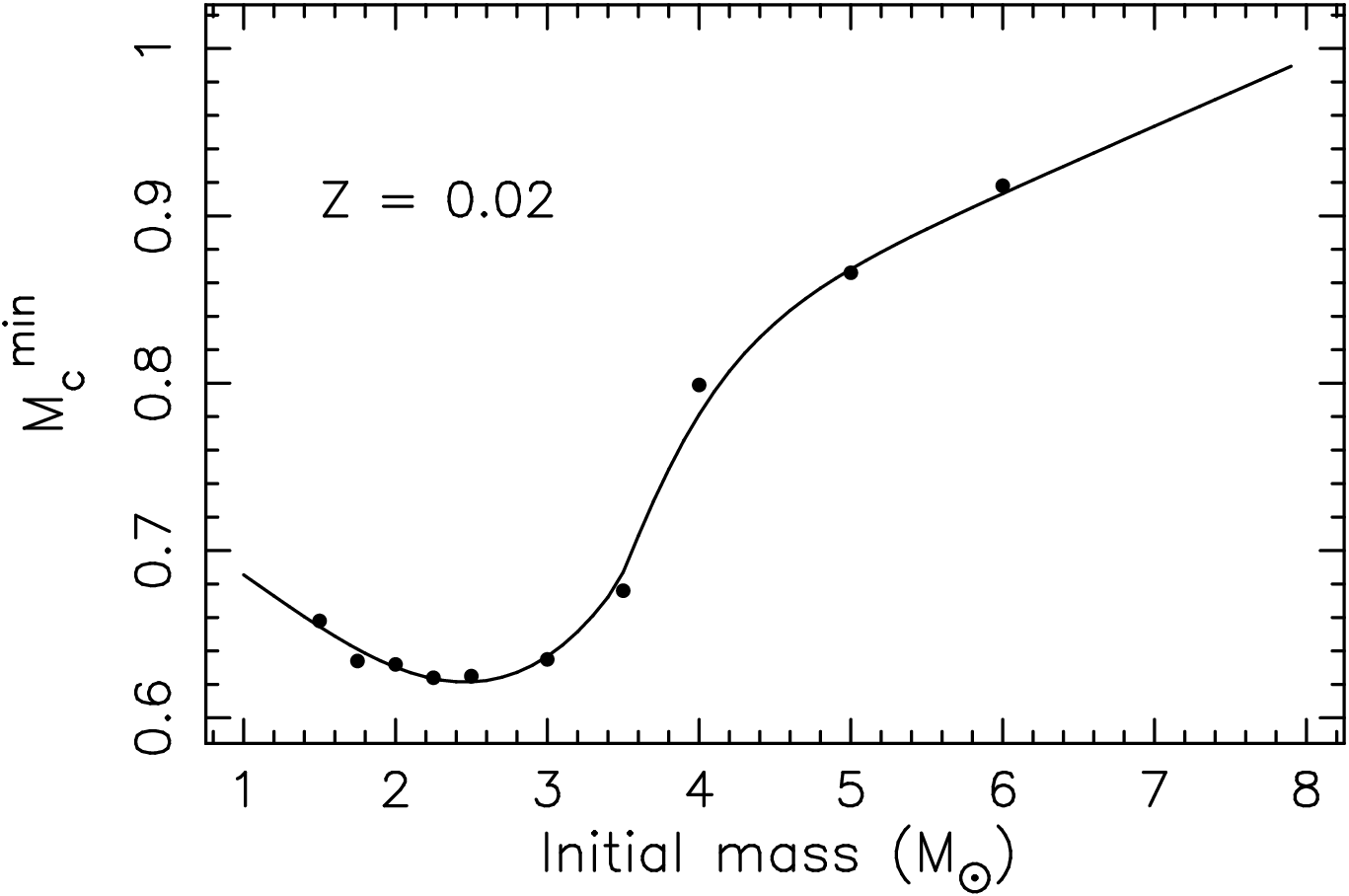,width=5.0cm,angle=270}  \\
   \psfig{file=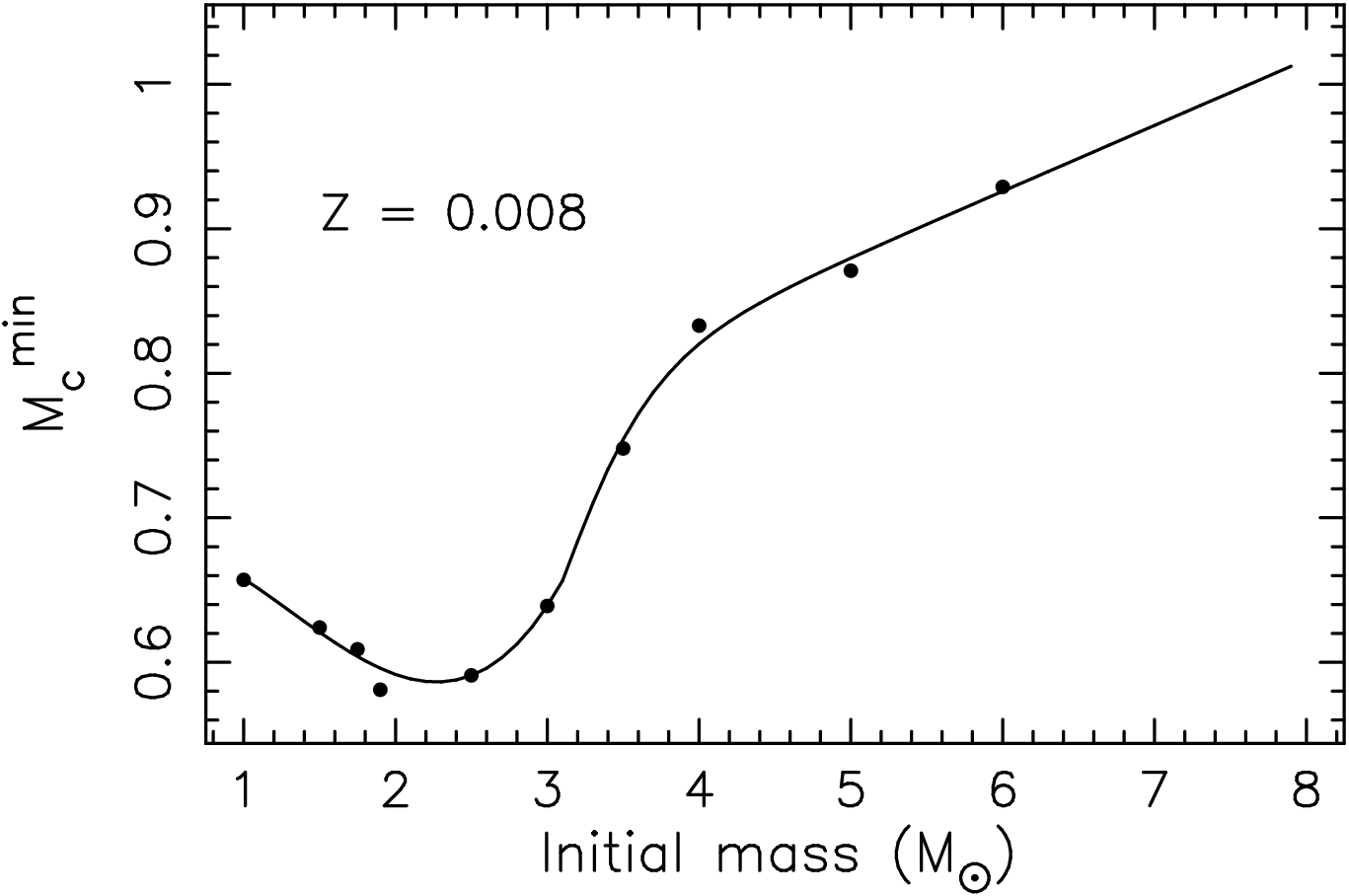,width=5.0cm,angle=270} \\
   \psfig{file=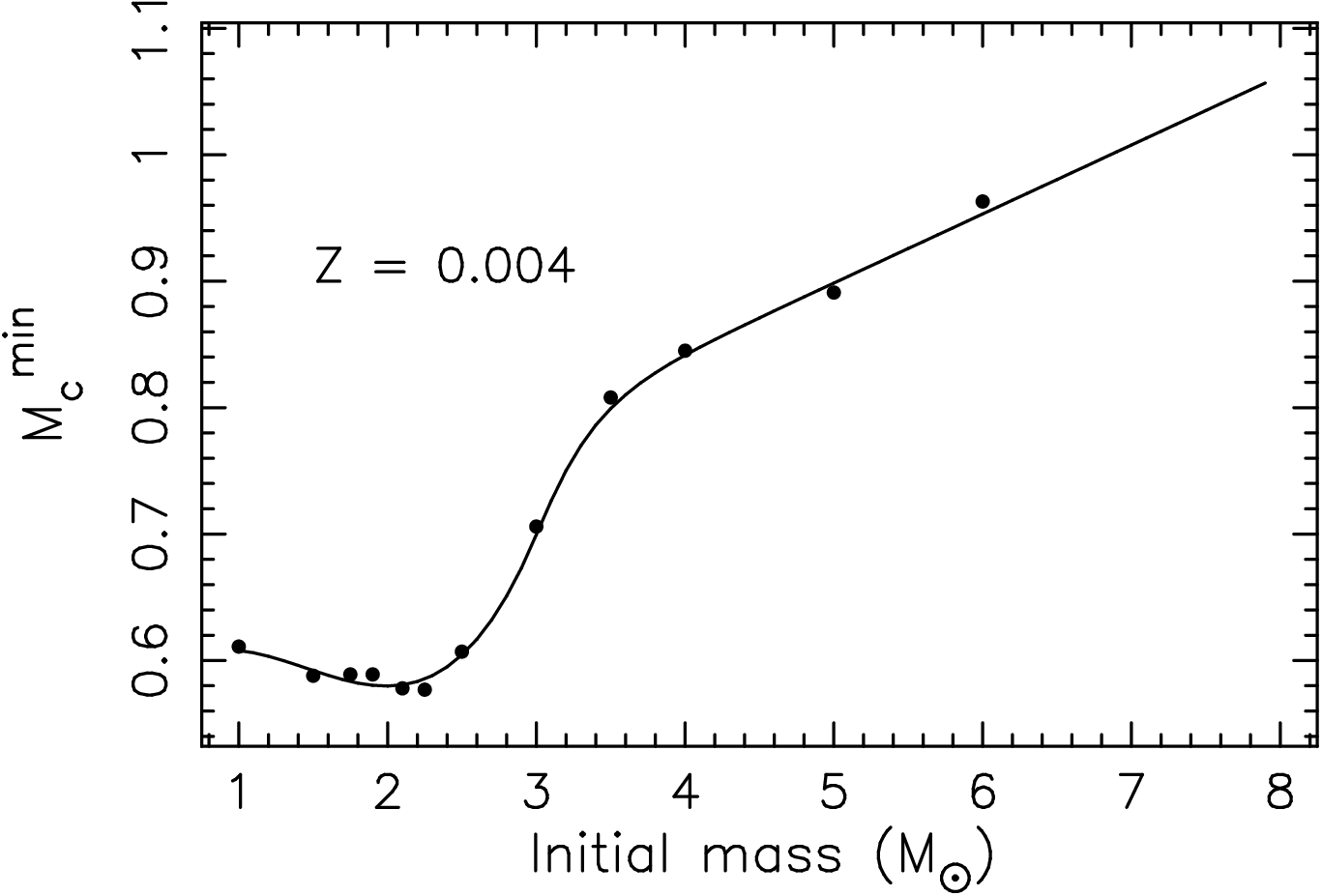, width=5.0cm,angle=270} \\
\end{tabular}
\caption{
The fit to $\Mcmin$ (solid line) for $Z=0.02$, $Z=0.008$
and $Z=0.004$, plotted with results (points) from the models without 
mass loss.
}
\label{fig:mcmin}
\end{figure}

From Tables 2, 3 and 4 it is clear that if $\McI > 0.7\Msun$, then
$\Mcmin$ has a value very close to $\McI$ (differing by less than
$0.005\Msun$). Hence it is justified to take $\Mcmin = \McI$ in this
case.
For lower masses generally $\Mcmin > \McI$, and
the behaviour of $\Mcmin$ is well approximated
by a third-order polynomial function.
Figure~\ref{fig:mcmin} shows the fits made to $\Mcmin$ as 
a function of total mass, for the case without mass loss.
The reader is referred to the Appendix for a full description of the
polynomial function and the coefficients.

The behaviour of $\Lmax$ is nearly linear at low $M$, rising 
steeply with mass until $M \sim 3\Msun$ before turning over 
and flattening out to be almost constant at high mass.
This behaviour is shown in 
Figure~\ref{fig:lmax} for the cases without mass loss.
The fits to $\Lmax$ in the figures were made 
with the function
\beqn
  \Lmax = \frac{b_{1} + b_{2} \MO + b_{3} \MO^{3}} {1 + b_{4} \MO^{3}}, \label{eq:rational}
\eeqn
where $b_{1}$, $b_{2}$, $b_{3}$, and $b_{4}$ are constants given in the Appendix. 
We note that as for $\McI$, the lines between $6\Msun$ and $8\Msun$ are
extrapolations from the fitting 
functions (valid to $6\Msun$) and may not reflect real model
behaviour, although one test calculation was made for $M=6.5$ and $Z=0.02$ 
and did agree with the fits presented here.

\begin{figure}
\begin{tabular}{c} 
    \psfig{file=fig15.eps,width=5.0cm,angle=270}  \\
   \psfig{file=fig16.eps,width=5.0cm,angle=270} \\
   \psfig{file=fig17.eps, width=5.0cm,angle=270} \\
\end{tabular}
\caption{
The fit to $\Lmax$ (solid line) for $Z=0.02$, $Z=0.008$
and $Z=0.004$, plotted with results (points) from the models without 
mass loss. }
\label{fig:lmax}
\end{figure}

\subsection{Dredge-up Parameter $\lambda$ as a function of time}

\begin{figure}
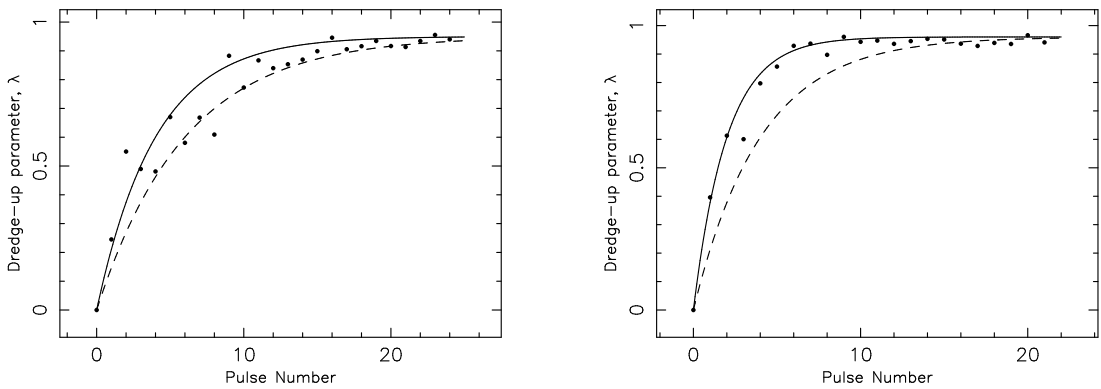

\begin{minipage}{0.5\textwidth}
\psfig{file=fig18.eps,width=5.0cm,angle=270}
\end{minipage}
\hspace{\fill}
\begin{minipage}{0.5\textwidth}
\psfig{file=fig19.eps,width=5.0cm,angle=270}
\end{minipage}
\hfill
\caption{({\it Left}) Fit to $\lambda$ from Equation~\ref{eq:expfit} with
$N_{\rm r} = 4$ (solid line) and $N_{\rm r} = 6$ (dashed line) 
for the $5\Msun$, $Z=0.02$ sequence without mass loss (points). 
We found the best fit to be $N_{\rm r} = 5$.
({\it Right}) Fit to $\lambda$ from Equation~\ref{eq:expfit} with 
$N_{\rm r}= 2$ (solid line)
and $N_{\rm r}= 4$ (dashed line) for the $5\Msun$, $Z=0.004$ sequence 
without mass loss.}
\label{fig:m5} 
\end{figure}

To accurately model the behaviour of the TDU 
we must include the increase of $\lambda$ over time.
For many of the low-mass models, $\lambda$ increases
slowly, only reaching $\Lmax$ after 8 or more thermal pulses.  
For the intermediate-mass models, $\lambda$ approaches 
$\Lmax$ asymptotically, reaching about $0.9 \Lmax$ in 
4 or more thermal pulses but it may not reach $\Lmax$ for many pulses.

To fit the behaviour of $\lambda$ in the models, 
we propose a simple method  shown in Figure~\ref{fig:m5}.
When $M_{\rm c} \geqslant \Mcmin$, $\lambda$ starts
increasing with pulse number, $N$, until $\lambda$ asymptotically 
reaches $\Lmax$ for large enough $N$.
Since our models gave little information on the decrease 
of $\lambda$ with decreasing
envelope mass, we suggest $\lambda = 0$ when 
$\Menv \leqslant M_{\rm env, crit}$, where 
$M_{\rm env, crit}$ is some critical value below which 
dredge-up does not occur. 
Low-mass models with dredge-up suggest that $M_{\rm env, crit} \lesssim 0.2$.

This behaviour can be modelled with the simple function:
\beqn
  \lambda (N) = \Lmax (1 - \exp^{-(N/N_{\rm r})}), \label{eq:expfit}
\eeqn
where $N$ is pulse number, measured from the first pulse where the 
core mass exceeds  $\Mcmin$.  $N_{\rm r}$ is a constant, 
determining how fast $\lambda$ reaches $\Lmax$.   
Due to the nature of the exponential function given by 
Equation~\ref{eq:expfit}, 
when $N > 8 N_{\rm r}$, Equation~\ref{eq:expfit} gives a value 
indistinguishable from $\Lmax$.
Table~\ref{tab:nr} lists the values of $N_{\rm r}$ which give 
the best fits to the models.

In finding an appropriate value of $N_{\rm r}$ for each model, 
we experimented with different 
values for each mass.  The increase in $\lambda$ observed in some 
models can be fitted by a range of $N_{\rm r}$ values, especially for 
models that exhibit a lot of scatter in their $\lambda$ values. 
For example, the $5\Msun$, $Z=0.02$ model without mass loss is one 
such case, where we find the range $4 \leqslant N_{\rm r} \leqslant 6$ 
gives a reasonable fit to the model as in Figure~\ref{fig:m5} (left).
The depth of dredge-up for the 
$5\Msun$, $Z=0.004$ model without mass loss is plotted 
against two fits from Equation~\ref{eq:expfit} in
Figure~\ref{fig:m5}.  We note that while the fit with 
$N_{\rm r} = 2$ approximates the model
behaviour best, the fit with $N_{\rm r} = 4$ is a good fit 
after 10 or more thermal pulses.
We also point out that the $5\Msun$, $Z=0.004$ model with mass loss 
experienced $\sim 80$ TPs,
so the first 10 or so pulses will be less important to the 
final composition of the star than
the first 10 pulses of a low-mass model that may only experiences 
30 or fewer pulses in total before the termination of the AGB phase.
For low-mass models with very small values of $\Lmax \lesssim 0.1$,  
Equation~\ref{eq:expfit} did not result in a good fit 
regardless of the $N_{\rm r}$ value used.  
We suggest setting $\lambda = \Lmax$ when $M_{\rm c} \geqslant \Mcmin$ for
these low-mass models.

From Table~\ref{tab:nr} we find a lot of variation in 
$N_{\rm r}$ with mass. Unfortunately, the variation
is not systematic and cannot be modelled with a simple function.
As we have argued above, the time dependence of $\lambda$ for 
low-mass stars is quite important as they have few TPs,
whilst more massive stars have many TPs so the first pulses are 
not so influential.
Therefore we suggest using a constant $N_{\rm r}$ value independent of
$M$ for a given $Z$, consistent with the low-mass models, e.g. 
$N_{\rm r} = 4$ for $Z=0.02$
and $N_{\rm r}=3$ for $Z=0.008$ and $Z=0.004$.

\begin{table}[ht]
\begin{center}
\caption{Table of $N_{\rm r}$ values for $Z=0.02$, $Z=0.008$ and $Z=0.004$.}
\vspace{2mm}
\begin{tabular}{@{}|c|c|c|c|@{}}  \hline
$\MO$    & $Z=0.02$ & $Z=0.008$ & $Z=0.004$ \\ \hline  
1.5      &  1       & 1        & 2  \\ 
1.75     &  3       & 3        & 3  \\
1.9      &  3  	    & 2        & 3  \\
2.25     &  4       & 3        & 3  \\
2.5      &  4       & 4        & 2  \\
3        & 3.5      & 4        & 1  \\
3.5      &  3       & 2        & 1   \\
4        &  2       & 2        & 1   \\
5        &  5       & 3        & 2   \\
6        &  4       & 3        & 3   \\ \hline 
\end{tabular} \label{tab:nr}
\end{center}
\end{table}

\section{Discussion}

\subsection{The Core Mass at the First Pulse}

The value of the core mass at the first thermal pulse is perhaps not
crucial to synthetic models, because it is the surface composition
changes caused by dredge-up that provide constraints on the
models. Hence it is $\Mcmin$ that is more important. Nevertheless,
comparisons with the CSLF in the Magellanic Clouds indicate that
detailed models overestimate $\Mcmin$ and it is useful to know $\McI$ which
is in principle the theoretical lower limit for $\Mcmin$. However,
$\McI$ may also be overestimated.

There are few parameterizations of this quantity in the literature.
Lattanzio (1989) gave a simple constant value for low mass stars,
and Renzini \& Voli (1981) gave a fit for more massive models.
These were used by Groenewegen \& de Jong (1993).
A more detailed fit was given by Wagenhuber \& Groenewegen (1998),
which was used by Marigo (2001). This
latter fit reproduces the shape very well. We have simply modified 
the coefficients as described in Section ~\ref{section:param-mci}
to provide a much better fit to the current results.

\subsection{Dredge-up: $\Mcmin$ and $\Lmax$}
Most synthetic calculations use constant $\Mcmin$ and constant 
$\lambda$. Groenewegen \& de Jong (1993) used the constant values
given by Lattanzio (1989) for $\Mcmin$, and then adjusted $\lambda$ to
try to fit the CSLF of the Magellanic Clouds. They found that $\Mcmin$ 
must also be decreased from the theoretical value, and they settled on 
$\Mcmin = 0.58$ and $\lambda=0.75$ to fit the observations.
A similar procedure was followed by Marigo et al (1996) and they found 
$\Mcmin = 0.58$ and $\lambda=0.65$. Note that Marigo (2001) now uses a 
more sophisticated algorithm for determining the onset of dredge-up,
as discussed in section~\ref{section:results}. 

The parameterisations we have given here
should be a significant improvement to the constant values used
for most synthetic
studies. In the discussion below we will compare our results with other
detailed evolutionary calculations.

From Figure~\ref{fig:lmax-mcmin} we find
$\Lmax$ to increase with decreasing $Z$ for a given mass, 
so we find low-mass models 
($\MO \leqslant 2\Msun,\,Z=0.008,0.004$ with mass loss) 
can became carbon stars with $\Lmax$ as high as $0.6$ for 
the $1.75\Msun,\,Z=0.004$ model.  
This effect is not so noticeable for higher mass stars ($M \gtrsim 4\Msun$),
where dredge-up quickly deepens with pulse number, and 
$\Lmax \approx 0.9$ for all compositions.

In comparison, Vassiliadis (1992), who  used a different version of the 
Mount Stromlo stellar evolution code (Wood \& Faulkner 1986, 1987) 
and older opacities (Huebner et al., 1977), only found dredge-up 
for $\MO \geqslant 2.5\Msun$ for LMC abundances and for 
$\MO \geqslant 2.0\Msun$ for SMC abundances. Clearly, the larger OPAL
opacities we use (Frost 1997) and the improved modelling of the 
TDU by Frost \& Lattanzio (1996) make a considerable difference.

Straniero et al. (1997), using the OPAL opacities find $\Lmax \approx 0.3$ for 
a solar composition $1.5\Msun$ model without mass loss. On the other
hand, we find $\Lmax \approx 0.05$,
substantially lower for the same mass and composition.
This is probably due to the difference in mixing-length parameter: we
used 1.75 and Straniero et al used the higher value of 2.2. 
A test calculation with a mixing-length parameter of 2.0 yielded 
$\Lmax = 0.2$.
We find deeper dredge-up than Straniero et al for the 
$3\Msun,\,Z=0.02$ model (without mass loss) 
with $\Lmax \approx 0.75$ where they find
$\Lmax \approx 0.46$.  These discrepancies must also be
related to the numerical
differences between the codes (Frost \& Lattanzio 1996; Lugaro 2001).

We find very similar values of $\lambda$ to Pols \& Tout (2001) 
for the $5\Msun$, $Z=0.02$ model. These authors
use a fully implicit method to solve the equations of 
stellar structure and convective mixing, 
and they find $\lambda$ to increase to $\approx 1.0$ in only six 
TPs while our models reach 
$\lambda \approx 0.95$ much more slowly (see Figure 6).

Herwig (2000) includes diffusive convective overshoot during all evolutionary
stages and on all convective boundaries on two solar composition models
of intermediate mass.
Without overshoot, no dredge-up is found for the 3$\Msun$ model. 
With overshoot, efficient dredge-up is found for both the 3 and 4$\Msun$ models,
where $\lambda \sim 1$ for the 3$\Msun$ model and  $\lambda > 1$ for the
4$\Msun$ model, which has the effect of decreasing the mass of the H-exhausted core
over time. Clearly, the inclusion of convective overshoot can 
substantially increase the amount of material dredged up from the intershell to the surface.
Langer et al. (1999) using a hydrodynamic stellar evolution code, 
model the effects of rotation on the structure and mixing of intermediate mass stars, 
also find some dredge-up in a 3$\Msun$ model of roughly solar composition. 

\subsection{The Carbon Star Luminosity Function}
The most common observation used to test the models is the
reproduction of observed CSLFs. 
We note that mass loss has the largest effect on the $Z=0.02$ 
models and we do not find any dredge-up for $M \leqslant 2\Msun$. 
It seems likely that our $Z=0.02$ models with mass loss 
cannot reproduce the low-mass end of the galactic
carbon star distribution, with
progenitor masses in the range $1-3\Msun$ (Wallerstein \& Knapp, 1998) as 
we find no dredge-up for models with mass-loss with 
$\MO \leqslant 2.0\Msun$.  The lowest mass solar composition model to become
a carbon star is the $3\Msun,\,Z=0.02$ model, which has $C/O \geqslant 1$
after 22 thermal pulses. We do note, however, that the 
galactic CSLF is very uncertain.

However, for LMC and SMC compositions, the CSLF is very well known
(see discussion in Groenewegen \& de Jong 1993).
It is a long standing problem that detailed evolutionary models fail to
match the observed CSLFs in the LMC and SMC (Iben 1981).
Although many of our  models with LMC and SMC compositions show enough 
dredge-up to turn them into carbon stars we expect that they will not
fit the low luminosity end of the
CSLF, because we find small values of $\lambda$ for 
$M \leqslant 1.5\Msun$, less than 
the value found from synthetic calculations of $\lambda \sim 0.5$ as the
required value to fit the CSLF. Also, we find larger $\Mcmin$ values for our
LMC and SMC low-mass models than the $0.58\Msun$ found from synthetic 
AGB calculations (Groenewegen \& de Jong, 1993; Marigo 1996, 1998a,b).

Within the context of synthetic models one usually modifies the
dredge-up law to ensure that agreement is reached. 
This usually means
decreasing $\Mcmin$ and increasing $\lambda$,
although this has previously been done crudely by altering constant
values for all masses (possibly with a composition
dependence)\footnote{Note that Marigo (1998a) adjusted her algorithm
via a reduction of $T_{\rm b}^{\rm dred}$ to 6.4 from the 6.7 found in
detailed models.}.
The models presented here 
show the variation with mass and composition of all dredge-up
parameters. This  has not been available previously. Although
modifications may 
be required, perhaps caused by our neglect of
overshoot (Herwig, 1997, 2000) or rotation (Langer et al., 1999),
we expect the dependence on mass and composition to be retained.

\section{Conclusions}
We have presented extensive evolutionary calculations covering a wide
range of masses and compositions, from the ZAMS to near the end of the 
AGB. Later papers will investigate 
nucleosynthesis and stellar yields, but 
in this paper we concerned ourselves with determining the dredge-up 
law operating in the detailed models. We have given parameterised fitting
formulae 
suitable for synthetic AGB calculations. As they stand, we expect that 
these will not fit the observed CSLFs in the LMC and SMC, a
long-standing problem. Some adjustments may be necessary, 
but  must be
consistent with the dependence on mass and composition 
as presented here. This may constrain the adjustments
and lead to a better understanding of where the detailed models
can be improved.

\section*{Appendix}

\subsection*{Coefficients for the fit to $\McI$}

The equations used by Wagenhuber \& Groenewegen (1998) to fit
$\McI$ are
\begin{eqnarray}
\McI &  =  & (-p_{1} (\MO - p_{2})^2 + p_{3})f + (p_{4}\MO + p_{5})(1-f), \\
   f &  =  & \left(1 + {\rm e}^{\frac{\MO - p_{6}}{p_{7}}} \right)^{-1}. \label{eq:mc1}
\end{eqnarray} 
Equations~8 and 9 are almost constant for stars with $\MO \leqslant 2.5\Msun$ 
and almost linear for stars that experience the second dredge-up (for masses
greater than about $4\Msun$).
The constant coefficients, $p_{1}$ through to $p_{7}$
that best fit our model results, are given in 
Table~\ref{tab:mc1fit}.

\begin{table}[ht]
\begin{center}
\caption{Coefficients for Equations~8 and 9: $p_{1}$, $p_{2}$, $p_{3}$, $p_{4}$, $p_{5}$, 
$p_{6}$ and $p_{7}$.}
\vspace{2mm}
\begin{tabular}{@{}|c|c|c|c|c|c|c|c|@{}}  \hline \hline
$Z$   & $p_{1}$ &  $p_{2}$ &  $p_{3}$ & $p_{4}$ & $p_{5}$ & $p_{6}$ & $p_{7}$ \\ \hline
0.02  & 0.038515 & 1.41379 & 0.555145 & 0.039781 & 0.675144 & 3.18432 & 0.368777 \\
0.008 & 0.057689 & 1.42199 & 0.548143 & 0.045534 & 0.652767 & 2.90693 & 0.287441 \\
0.004 & 0.040538 & 1.54656 & 0.550076 & 0.054539 & 0.625886 & 2.78478 & 0.227620 \\ \hline
\end{tabular} \label{tab:mc1fit}
\end{center} 
\end{table} 
 
\subsection*{Coefficients for the fits to $\Mcmin$ and $\Lmax$}

Let $M_{\rm sdu}$ be the minimum mass, at a given composition, which
experiences the second dredge-up. Hence from our models
$M_{\rm sdu} = 4\Msun$ for $Z=0.02$, $3.8\Msun$ for $Z=0.008$ and
$3.5\Msun$ for $Z=0.004$. 
For masses $\MO < M_{\rm sdu}$, our results for $\Mcmin$ are fitted to
a cubic polynomial,
\beqn
 \Mcmin = a_{1} + a_{2} \MO + a_{3} \MO^{2} + a_{4} \MO^{3}  \label{eq:cubic}
\eeqn
where $a_{1}$, $a_{2}$, $a_{3}$ and $a_{4}$ are constants that depend on 
$Z$ and are given in Table~\ref{tab:mcmin} and $\MO$ is the initial mass of 
the star (in solar units).

For cases where $\MO \gtrsim M_{\rm sdu} - 0.5\Msun$ 
we find that $\Mcmin > 0.70\Msun$, 
and we can set $\Mcmin = \McI$ consistent
with our model results. Since Eq.~\ref{eq:cubic} diverges for large masses,
in practice we recommend calculating $\Mcmin$ by
the following procedure:
\beqn
 \Mcmin = \max(\McI, \min(0.7\Msun, \Mcmin{^\ast}))  \label{eq:mcmin}
\eeqn
where $\Mcmin{^\ast}$ is given by Eq.~\ref{eq:cubic}. This ensures that always
$\Mcmin \geqslant \McI$ as required, while  $\Mcmin = \McI$ if 
$\McI > 0.7\Msun$.

\begin{table}[ht]
\begin{center}
\caption{$a_{1}$, $a_{2}$, $a_{3}$ and $a_{4}$ for Equation~\ref{eq:cubic}.}
\vspace{2mm}
\begin{tabular}{@{}|c||c|c|c|c|@{}}  \hline
$Z$   &  \multicolumn{4}{c|}{$\Mcmin$}  \\ \hline \hline
      & $a_{1}$ & $a_{2}$ & $a_{3}$ & $a_{4}$  \\ \hline
0.02  & 0.732759 & -0.0202898 & -0.0385818 & 0.0115593 \\
0.008 & 0.672660 &  0.0657372 & -0.1080931 & 0.0274832 \\
0.004 & 0.516045 &  0.2411016 & -0.1938891 & 0.0446382 \\ \hline  
\end{tabular} \label{tab:mcmin}
\end{center}
\end{table}

We fit $\Lmax$ with a rational polynomial of the type given in Equation~\ref{eq:rational}.
The constants, $b_{1}$, $b_{2}$, $b_{3}$ and $b_{4}$ for $Z=0.02$, $Z=0.008$ and $Z=0.004$ 
are given in Table~\ref{tab:lmax}.
For $Z=0.02$, we only fit $\Lmax$ and $\Mcmin$ down to $1.5\Msun$ and as a consequence
the fit to $\Lmax$ goes negative for masses below this. For $Z=0.008$ and $Z=0.004$, we
fit $\Lmax$ and $\Mcmin$ down to $1\Msun$. 
Therefore, if Equation~\ref{eq:rational} yields a negative value $\Lmax$ should 
be set to zero.

\begin{table}[ht]
\begin{center}
\caption{$b_{1}$, $b_{2}$, $b_{3}$ \& $b_{4}$ for Equation~\ref{eq:rational} for $\Lmax$}
\vspace{2mm}
\begin{tabular}{@{}|c|c|c|c|c|c|c|c|c|@{}}  \hline
$Z$  &  \multicolumn{4}{c|}{$\Lmax$}  \\ \hline \hline
& $b_{1}$ & $b_{2}$ & $b_{3}$ & $b_{4}$ \\ \hline
0.02  & -1.17696 & 0.76262 &  0.026028 & 0.041019 \\
0.008 & -0.609465 & 0.55430 & 0.056878 & 0.069227 \\
0.004 & -0.764199 & 0.70859 & 0.058833 & 0.075921 \\ \hline
\end{tabular} \label{tab:lmax}
\end{center}
\end{table}

It is possible to linearly interpolate between the coefficients in $Z$
to find fits for intermediate metallicities. This may not reflect real model behaviour
but the functions are well behaved.  Note that interpolating between the coefficients
of Equation~\ref{eq:rational} in the range $0.02 < Z < 0.008$ will result in negative
values of $\Lmax$ between $1 \leqslant \MO (\Msun) \leqslant 1.5$.  Again we suggest
setting $\Lmax = 0$ when this happens.

\section*{Acknowledgments}

AIK would like to acknowledge the assistance of a Monash Graduate Scholarship for 
support and the Victorian Partnership for Advanced Computing for computational 
time and support. We also thank the anonymous referees for helping to 
improve the clarity of the paper.


\section*{References}






\reference Boothroyd,~A.~I. \& Sackmann,~I.~-J., 1988, ApJ, 328, 671
\reference Busso,~M., Gallino,~R. \& Wasserburg,~G., 1999, ARA\&A, 37, 329
\reference Frost,~C.~A., 1997, Ph.D. Thesis, Monash University
\reference Frost,~C.~A. \& Lattanzio,~J.~C., 1995, in Stellar Evolution: What Should Be Done? 
\reference Frost,~C.~A. \& Lattanzio,~J.~C., 1996, ApJ, 344, L25
\reference Groenewegen,~M.~A.~T \& de~Jong,~T., 1993, A\&A, 267, 410
\reference Herwig,~F., 2000, A\&A, 360, 952
\reference Herwig,~F., Bl\"{o}cker,~T., Sch\"{o}nberner,~D., El Eid,~M., 1997, A\&A, 324, L81
\reference Huebner,~W.~F., Merts,~A.~L., Magee,~N.~H.~Jr. \& Argo,~M.~F., 1977, 
Astrophysical Opacity Library, Los Alamos Scientific Laboratory, LA-6760-M
\reference Hurley,~J.~R., Tout,~C.~A., Pols,~O.~R., 2002, MNRAS, 329, 897
\reference Iben~Jr.,~I., 1981, ApJ, 246, 278
\reference Iben~Jr.,~I., 1991, ed. G.~Michaud, A.~V.~Tutukov, in 
Evolution of Stars: The Photospheric
Abundance Connection, (Dordrecht, Kluwer Academic Publishers), 257
\reference Iglesias,~C.~A. \& Rogers,~F.~J., 1996, ApJ, 464, 943
\reference Langer~N., Heger~A., Wellstein~S., Herwig~F., 1999, A\&A, 346, L37
\reference Lugaro,~M.~A., 2001, Ph.D. Thesis, Monash University
\reference Marigo,~P., Bressan,~A. \& Chiosi,~C., 1996, A\&A, 313, 545
\reference Marigo,~P., 1998a, in Asymptotic Giant Branch Stars, ed. T.~Le Bertre,
 A.~L\`{e}bre, C.~Waelkens, (Astronomical Society of the Pacific), 53
\reference Marigo,~P., Bressan,~A. \& Chiosi,~C., 1998b, A\&A, 331, 564
\reference Marigo,~P., 2001, A\&A, 370, 194
\reference Mowlavi,~N., 1999, A\&A, 344, 617
\reference Pols,~O.~R., Tout,~C.~A., 2001, in Salting the Early Soup: Trace Nuclei from 
Stars to the Solar System, ed. M~.Busso \& R.~Gallino, Mem. S.A.It., v72 No. 2, in press.
\reference Straniero,~O., Chieffi,~A., Limongi,~M., Busso,~M., Gallino,~R. 
\& Arlandini,~C., 1997, ApJ, 478, 332
\reference Reimers,~D., 1975, in Problems in Stellar Atmospheres and Envelopes, 
ed. B.~Baschek, W.~H. Kegel, G.~Traving, (New York: Springer-Verlag), 229
\reference Wagenhuber,~J. \& Groenwegen,~M.~A.~T., 1998, A\&A, 340, 183
\reference Wallerstein,~G. \& Knapp,~G.~R., 1998, ARA\&A, 36, 369
\reference Wood,~P.~R. \& Faulkner,~D.~J, 1986, ApJ, 307, 659
\reference Wood,~P.~R. \& Faulkner,~D.~J, 1987, PASA, 7, 75
\reference Wood,~P.~R. \& Zarro,~D.~M., 1981, ApJ, 248, 311
\reference van~den~Hoek,~L.~B. \& Groenewegen,~M.~A.~T, 1997, A\&ASS, 123, 305
\reference Vassiliadis,~E., 1992, Ph.D. Thesis, Australian National University
\reference Vassiliadis,~E. \& Wood,~P.~R., 1993, ApJ, 413, 641

\end{document}